\documentclass{biometrika}

\usepackage[plain,noend]{algorithm2e}

\usepackage{natbib,color,bm,amsmath,amssymb,graphicx,float,booktabs,subfig,multirow,threeparttable,textcomp,times}
\usepackage{tikz}
  \usetikzlibrary{positioning,shapes.geometric}

\makeatletter
\renewcommand{\algocf@captiontext}[2]{#1\algocf@typo. \AlCapFnt{}#2} 
\def\@algocf@capt@plain{top}
\renewcommand{\algocf@makecaption}[2]{%
  \addtolength{\hsize}{\algomargin}%
  \sbox\@tempboxa{\algocf@captiontext{#1}{#2}}%
  \ifdim\wd\@tempboxa >\hsize
    \hskip .5\algomargin%
    \parbox[t]{\hsize}{\algocf@captiontext{#1}{#2}}
  \else%
    \global\@minipagefalse%
    \hbox to\hsize{\box\@tempboxa}
  \fi%
  \addtolength{\hsize}{-\algomargin}%
}
\makeatother

\def\Bka{{\it Biometrika}}

\begin{document}

\newcommand{\red}{\color{red}}
\newcommand{\blue}{\color{blue}}
\newcommand{\mage}{\color{magenta}}

\def\logit{\text{\rm logit }}
\def\expit{\text{\rm expit }}
\def\pr{\text{\rm pr}}

\newcommand\independent{\protect\mathpalette{\protect\independenT}{\perp}}
\def\independenT#1#2{\mathrel{\rlap{$#1#2$}\mkern2mu{#1#2}}}
\makeatletter
\newcommand*{\ind}{%
\mathbin{%
\mathpalette{\@ind}{}%
}%
}
\newcommand*{\nind}{%
\mathbin{
\mathpalette{\@ind}{\not}
}%
}
\newcommand*{\@ind}[2]{%
\sbox0{$#1\perp\m@th$}
\sbox2{$#1=$}
\sbox4{$#1\vcenter{}$}
\rlap{\copy0}
\dimen@=\dimexpr\ht2-\ht4-.2pt\relax
\kern\dimen@
{#2}%
\kern\dimen@
\copy0 
} 
\makeatother

\jname{Biometrika}
\jyear{2017}
\jvol{}
\jnum{}
\copyrightinfo{\Copyright\ 2012 Biometrika Trust\goodbreak {\em Printed in Great Britain}}
%

\markboth{Wang Miao et al.}{Miscellanea}
%
%
%
%

\title{Identifying  Causal Effects With Proxy Variables of an Unmeasured Confounder}

\author{Wang Miao}
\affil{Guanghua School of Management, Peking University, 5 Summer Palace Road, Haidian District,    Beijing 100871, P.R.C. \email{mwfy@pku.edu.cn}}

\author{ Zhi Geng}
\affil{School of Mathematical Sciences, Peking University,   5 Summer Palace Road, Haidian District, Beijing 100871, P.R.C. \email{zhigeng@pku.edu.cn}}

\author{\and Eric Tchetgen Tchetgen}
\affil{Department of Biostatistics, Harvard University, 677 Huntington Avenue, Boston,  Massachusetts 02115, U.S.A. \email{etchetge@hsph.harvard.edu}}

\maketitle

\begin{abstract}
We consider a causal  effect  that is confounded by an unobserved  variable, but with observed proxy variables  of the  confounder. We show that, with at least two independent   proxy variables  satisfying  a certain rank condition,   the causal effect is nonparametrically identified, even if the  measurement error mechanism, i.e., the conditional distribution of the proxies given the confounder,  may  not be identified. Our result generalizes the identification strategy of \cite{kuroki2014measurement} that rests on identification of the measurement error mechanism.
When  only one proxy for the confounder is available, or the required rank condition is not met, we develop a strategy to test the null hypothesis of no causal effect.
\end{abstract}

\begin{keywords}
	Confounder;  Identification; Measurement error; Negative control; Proxy.
\end{keywords}
\section{Introduction}
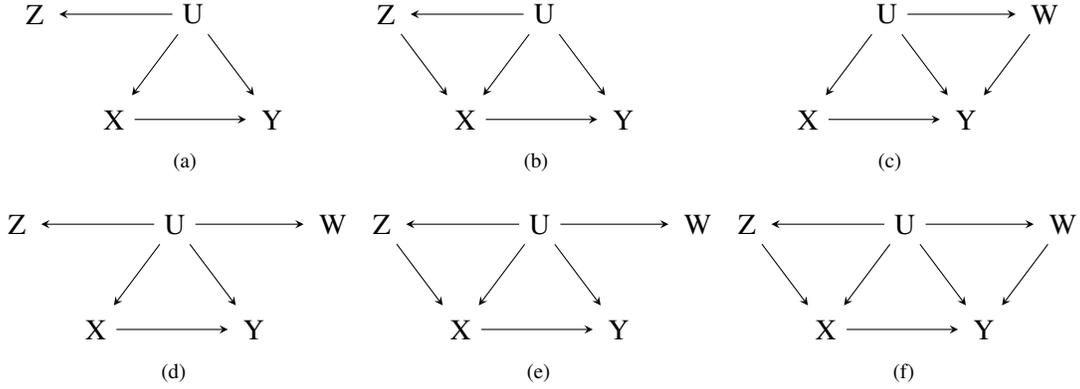
\begin{figure}[h]
	\centering
	\subfloat[]{
	\begin{tikzpicture}[scale=0.7,
	->,
	shorten >=2pt,
	>=stealth,
	node distance=1cm,
	pil/.style={
		->,
		thick,
		shorten =2pt,}
	]
	\node (X) at (1.5,-2) {X};
	\node (U) at (3,0) {U};
	\node (Z) at (0,0) {Z};
	\node (Y) at (4.5,-2) {Y};
	\node (W) at (6,0) {};

	\foreach \from/\to in {X/Y,U/Y,U/X,U/Z}
	\draw (\from) -- (\to);

	\end{tikzpicture}}
\subfloat[]{
	\begin{tikzpicture}[scale=0.7,
	->,
	shorten >=2pt,
	>=stealth,
	node distance=1cm,
	pil/.style={
		->,
		thick,
		shorten =2pt,}
	]
	\node (X) at (1.5,-2) {X};
	\node (U) at (3,0) {U};
	\node (Z) at (0,0) {Z};
	\node (Y) at (4.5,-2) {Y};
	\node (W) at (6,0) {};

	\foreach \from/\to in {X/Y,U/Y,U/X,U/Z,Z/X}
	\draw (\from) -- (\to);

	\end{tikzpicture}}
\subfloat[]{
		\begin{tikzpicture}[scale=0.7,
		->,
		shorten >=2pt,
		>=stealth,
		node distance=1cm,
		pil/.style={
			->,
			thick,
			shorten =2pt,}
		]
		\node (X) at (1.5,-2) {X};
		\node (U) at (3,0) {U};
		\node (W) at (6,0) {W};
		\node (Y) at (4.5,-2) {Y};
		\node (Z) at (0,0) {};

		\foreach \from/\to in {X/Y,U/Y,U/X,U/W,W/Y}
		\draw (\from) -- (\to);

		\end{tikzpicture}}\\
	\subfloat[]{
		\begin{tikzpicture}[scale=0.7,
		->,
		shorten >=2pt,
		>=stealth,
		node distance=1cm,
		pil/.style={
			->,
			thick,
			shorten =2pt,}
		]
		\node (X) at (1.5,-2) {X};
		\node (U) at (3,0) {U};
		\node (W) at (6,0) {W};
		\node (Y) at (4.5,-2) {Y};
		\node (Z) at (0,0) {Z};

		\foreach \from/\to in {X/Y,U/Y,U/X,U/W,U/Z}
		\draw (\from) -- (\to);

		\end{tikzpicture}}
\subfloat[]{
	\begin{tikzpicture}[scale=0.7,
	->,
	shorten >=2pt,
	>=stealth,
	node distance=1cm,
	pil/.style={
		->,
		thick,
		shorten =2pt,}
	]
	\node (X) at (1.5,-2) {X};
	\node (U) at (3,0) {U};
	\node (W) at (6,0) {W};
	\node (Y) at (4.5,-2) {Y};
	\node (Z) at (0,0) {Z};

	\foreach \from/\to in {X/Y,U/Y,U/X,U/W,U/Z,Z/X}
	\draw (\from) -- (\to);
	
	\end{tikzpicture}}
\subfloat[]{
	\begin{tikzpicture}[scale=0.7,
	->,
	shorten >=2pt,
	>=stealth,
	node distance=1cm,
	pil/.style={
		->,
		thick,
		shorten =2pt,}
	]
	\node (X) at (1.5,-2) {X};
	\node (U) at (3,0) {U};
	\node (W) at (6,0) {W};
	\node (Y) at (4.5,-2) {Y};
	\node (Z) at (0,0) {Z};

	\foreach \from/\to in {X/Y,U/Y,U/X,U/W,U/Z,W/Y,Z/X}
	\draw (\from) -- (\to);
	
	\end{tikzpicture}}
	\caption{Causal diagrams with confounder proxies.}\label{fig:1}
\end{figure}

Unmeasured confounding is a crucial problem in observational studies.  Simpson's (1951) \nocite{simpson1951interpretation} paradox 
is an elegant illustration of the type of bias that may arise in causal inference subject to unmeasured confounding.
Sometimes an analyst may have access to one or more proxies for the unobserved confounder, for example, a mismeasured version of the confounder. In this case, it may seem natural to directly adjust for the available proxies in order  to reduce bias due to unobserved confounding \citep{greenland1980,greenland1996basic,carroll2006measurement,ogburn2013bias,kuroki2014measurement}. 
\cite{greenland1980} suggested  that   standard adjustment  of a  binary nondifferential proxy that is independent of the treatment and the outcome after conditioning on the confounder  generally  reduces  bias due to confounding; for a polytomous confounder, however, certain  monotonicity assumptions appear  indispensable to guarantee such bias attenuation \citep{ogburn2013bias}. But even when the monotonicity assumptions  are met, the  approach of \cite{greenland1980} and \cite{ogburn2013bias}  cannot completely eliminate the confounding bias. 
\cite{greenland2008bias} developed a matrix adjustment method that can 
completely account for unobserved confounding, but it requires external information on  the error mechanism, i.e., the  conditional distribution of the  proxy given the  confounder, and therefore cannot be applied in routine situations where the error mechanism  is unknown.
Fortunately, when multiple proxies for the confounder are available, 
as shown by \cite{kuroki2014measurement}, one can sometimes identify the error mechanism and  thus the causal effect  without external information. \cite{kuroki2014measurement} studied identification of the causal effect with  two independent proxies  in the context of graph-based models, where identification means that the causal effect  can be determined uniquely from the joint distribution of observed variables. Figure \ref{fig:1} presents several plausible  causal diagrams with proxies for the confounder, where $X$ and $Y$ denote the treatment and the outcome respectively, $U$ denotes the confounder that is not observed, but  proxies  $Z$ and $W$ of $U$ may be  available. 
Model  (a) corresponds to a single nondifferential proxy \citep{carroll2006measurement,greenland2008bias}, and (b)--(c) allow the  treatment and the outcome to depend on the proxy, respectively; models (d)--(f) depict  situations where  two proxies are independent conditional on the true confounder.  For graph-theoretic terminology, we refer readers to \cite{pearl2009causality}. Table \ref{tbl:1} presents the corresponding conditional independencies for the diagrams in Figure \ref{fig:1}.
Practical examples for such diagrams can be found in \cite{carroll2006measurement}, \cite{greenland2008bias} and \cite{kuroki2014measurement}.
Using the  ${\rm do}(x)$ operator of \cite{pearl2009causality}, the  causal effect of $X$ on $Y$ is 
\[\pr\{y\mid {\rm do}(x)\}=\sum_u \pr (y\mid x,u)\pr (u),\] 
where  $\pr$  stands for the probability mass function of a discrete variable or the probability density function for a  continuous variable.
For (d)--(e), \cite{kuroki2014measurement} establish sufficient conditions for nonparametric  identification of   $\pr(w\mid u)$, which suffices  to identify  $\pr\{y\mid \text{do}(x)\}$ by  applying the matrix adjustment technique  of \cite{greenland2008bias}.

Model (f) is more general than (d)--(e). Only under a joint normal model, \cite{kuroki2014measurement} established identification of the causal effect for (f). 
The  nonparametric identification method of \cite{kuroki2014measurement}  depends on   identification of  $\pr(w\mid u)$, and  does not  apply to  model (f), because   in general,   $\pr(w\mid u)$ is  not identifiable in (f).  Nonparametric  identification of the causal effect for  model (f)  is  not yet available.
Below,  we propose a novel strategy to nonparametrically  identify the causal effect for  model (f)  without identifying  $\pr(w\mid u)$.  We consider a categorical confounder in Section 2, and then   we generalize the  results to the  continuous case in Section 3.
The required condition is  weaker  than that  of \cite{kuroki2014measurement}.
Moreover, when only one proxy is available or the proposed identification condition   is not met, 
we   establish that it is nonetheless sometimes possible to obtain a valid empirical test of the null hypothesis of no causal effect.

\begin{table}[h]
\centering
\caption{Conditional independencies of causal diagrams}\label{tbl:1}
\begin{tabular}{ll}
 (a) $Z\ind (X,Y)\mid U$ & (b) $Z\ind Y\mid (U,X)$\\
 (c) $W\ind X\mid U$ & (d) $W\ind (Z,X,Y)\mid U$, $Z\ind (X,Y)\mid U$\\
 (e) $W\ind (Z,X,Y)\mid U$, $Z\ind Y\mid (U,X)$  & (f) $W\ind (Z,X)\mid U$, $Z\ind Y\mid (U,X)$\\
\end{tabular}

\end{table}

\section{Identification with a categorical confounder}
As (d) and (e)  can be viewed as special cases of  model (f)  with  $W\ind Y\mid U$, we focus on identification of the causal effect for  (f).
Suppose  $W,Z$ and $U$ are discrete variables, each with  $k$ categories. For notational convenience, we use 
$P(W\mid u)=\{\pr(w_1\mid u),\ldots,\pr(w_k\mid u)\}^{\rm T}$, $P(w\mid U)=\{\pr(w\mid u_1),\ldots,\pr(w\mid u_k)\}$  and  $P(W\mid U)=\{P(W\mid u_1),\ldots,P(W\mid u_k)\}$ 
to denote a column vector,  a row vector  and  a matrix  that consist of conditional probabilities $\pr(w\mid u)$, respectively.
For other variables, vectors and matrices consisting of conditional probabilities  are analogously defined: $P(U\mid z,x)=\{\pr(u_1\mid z,x),\ldots,\pr(u_k\mid z,x)\}^{\rm T}$, 
$P(U\mid Z,x)=\{P(U\mid z_1,x),\ldots,P(U\mid z_k,x)\}$, and  $P(y\mid Z,x)=\{\pr(y\mid z_1,x),\ldots,\pr(y\mid z_k,x)\}$.
Model  (f) implies  that $W\ind (Z,X)\mid U$ and $Z\ind Y\mid (U,X)$, so
\begin{equation}
  P(W\mid Z, x)  =    P(W\mid U)  P(U\mid Z, x),\label{eq:wzt}
\end{equation}
\begin{equation}
  P(y\mid Z, x)  =    P(y\mid U, x)  P(U\mid Z, x)\label{eq:yzt}.
\end{equation}
Based on \eqref{eq:wzt} and \eqref{eq:yzt},
we  identify  the causal effect under the condition
\begin{enumerate}
	\item[(i)] $  P(W\mid  Z,  x)$ is  invertible for all  $x$.
\end{enumerate}
This is weaker than the condition  of \cite{kuroki2014measurement}, which  requires invertibility of both $P(W\mid Z,x)$ and $P(y,W\mid Z,x)$.
Condition (i) is  equivalent to requiring that  both  $P(W\mid U)$ and $P(U\mid Z,x)$ are invertible   \citep[Corollary 5.4]{banerjee2014}, which implies that   both   $W$ and $Z$ are associated with the confounder $U$.
Condition (i) involves only $W,Z$ and $X$, and thus can be verified empirically.   For the binary case,  (i)    holds if both $Z$ and $W$ are  correlated within each level of $X$.
Under (i),  \eqref{eq:wzt}--\eqref{eq:yzt} imply
\begin{align}
P(U\mid Z, x)  & =  P(W\mid U) ^{-1} P(W\mid Z,x), \nonumber\\
P(y\mid Z, x)  & =    P(y\mid U, x)  P(W\mid U)^{-1} P(W\mid Z, x), \label{eq:yzt2}
\end{align}
and thus
\begin{align}\label{eq:yux}
P(y\mid U,x)=    P(y\mid Z,x)   P(W\mid Z, x)^{-1}  P(W\mid U).
\end{align}
From \eqref{eq:yux}, identification of $P(y\mid U,x)$ does depend on  the error mechanism  $P(W\mid U)$. 
Letting $  P(U)=\{\pr (u_1),\ldots, \pr(u_k)\}^{\rm T}$  with $P(W)$ analogously defined,  we have $P(W)=P(W\mid U)P(U)$.  
Multiplying $P(U)$ on both sides of \eqref{eq:yux}, we obtain
\begin{equation}\label{eq:ce1}
\pr\{y\mid {\rm do}(x)\} =    P(y\mid Z, x)   P(W\mid Z, x)^{-1}  P(W).
\end{equation}
As a result,  identification of $\pr\{y\mid\text{do}(x)\}$ does not depend on    $P(W\mid U)$!
In contrast,  the approach of \cite{kuroki2014measurement}  requires  identification of   $P(W\mid U)$, which rests on the assumption $W\ind Y\mid U$.    When   $W\nind Y\mid U$, i.e., for model (f), however, $P(W\mid U)$ is  in general not identified, and therefore, their approach fails even though  the causal effect may still be identified by our formula \eqref{eq:ce1}. We illustrate this  in the Supplementary Material.

For the binary case,   the right hand side of \eqref{eq:ce1} reduces to
\[\frac{\{\pr(w_1)-\pr(w_1\mid z_2,x)\}\pr(y\mid z_1,x)}{\pr(w_1\mid z_1,x)-\pr(w_1\mid z_2,x)}+\frac{\{\pr(w_1\mid z_1,x)-\pr(w_1)\}\pr(y\mid z_2,x)}{\pr(w_1\mid z_1,x)-\pr(w_1\mid z_2,x)},\]
which is a weighted  average of $\pr(y\mid z_i,x), i=1,2$.
It can be viewed as  a modified version  of the adjustment formula $\sum_{i=1}^{2}\pr(y\mid z_i,x)\pr(z_i)$  suggested by \cite{greenland1980} and \cite{ogburn2013bias}.
But their approach can  incorporate only one proxy and is  biased for $\pr\{y\mid{\rm do}(x)\}$ due to  confounding.  As a second  proxy $W$ is available, instead of the weight $\pr(z)$,  we use a modified weight  that can eliminate the bias due to imperfect adjustment of $Z$.

If $W$ and $Z$  have more  categories than $U$,  the causal effect is  identifiable as long as $P(W\mid Z,x)$ has rank $k$. Identification for this case is achieved by using corresponding  coarsening variables $W'$ and $Z'$ with $k$ categories  such that $P(W'\mid Z',x)$ is invertible.

\section{Identification with a continuous confounder}
In empirical studies,  unobserved confounders may  sometimes be continuous. 
Under  model (f), we  generalize  \eqref{eq:ce1} to the continuous case   by assuming  the following completeness condition: for all square-integrable function $g$ and for any  $x$,
\begin{enumerate}
\item [(ii)]  $E\{g(u)\mid z,x\}=0$ almost surely if and only if $g(u)=0$ almost surely;
\item [(iii)]  $E\{g(z)\mid w,x\}=0$ almost surely if and only if $g(z)=0$ almost surely.
\end{enumerate}
Conditions (ii)--(iii) can accommodate both categorical and continuous confounders. For a categorical  confounder with categorical proxy variables, (ii)--(iii)    is equivalent to (i). 
For a continuous confounder,  we suppose that both $Z$ and $W$   are continuous.  Many commonly-used parametric and semiparametric models  such as   exponential families \citep{newey2003instrumental} and  location-scale families \citep{hu2017nonparametric} satisfy  the completeness condition.
For nonparametric  regression models, results of  \cite{d2011completeness} and \cite{darolles2011nonparametric}    can be used  to justify the completeness condition, although they  focused on   instrumental variable estimation.
For a review and examples of completeness, see \cite{chen2014local}, \cite{andrews2017examples} and the references therein.

Letting  $f$  denote the probability density  function of a continuous  variable, instead of the matrix form   \eqref{eq:yzt2} for the categorical case, identification for the continuous case involves the solution  $h(w,x,y)$ to the  following  integral equation: for any $(x,y)$ and for all $z$,
\begin{equation}\label{eq:intg}
\pr(y\mid z,x)=\int_{-\infty}^{+\infty}h(w,x,y)f(w\mid z,x)dw.
\end{equation} 
Equation \eqref{eq:intg} is a Fredholm integral equation of the first kind. A conventional  approach to study its solution  is the  singular value decomposition  \citep[Theorem 2.41]{carrasco2007}. 
We show existence of its solution  under    (iii) together with   regularity conditions (v)--(vii) in Proposition \ref{prop:1} in the Appendix.
The solution to \eqref{eq:intg} may not be unique, but the causal effect can still be identified.

\begin{theorem}\label{thm:idn}
Assuming    model (f)  and condition (ii),  for any solution $h(w,x,y)$ to \eqref{eq:intg}, 
\begin{eqnarray}
\pr(y\mid u,x)  &=&  \int_{-\infty}^{+\infty} h(w,x,y)f(w\mid u)dw, \label{eq:cce}\\
\pr\{y\mid {\rm do}(x)\} & =&  \int_{-\infty}^{+\infty} h(w,x,y)f(w)dw. \label{eq:ce2}
\end{eqnarray}
\end{theorem}
From the theorem, one can identify the causal effect by first solving \eqref{eq:intg} and then applying  \eqref{eq:ce2} with $\pr(y\mid z,x)$, $f(w\mid z,x)$ and  $f(w)$ obtained from observed variables.  
While Theorem \ref{thm:idn} provides a formal basis for nonparametric inference, specific details on how such inferences can be obtained are  beyond the scope of this paper.
Nevertheless,  we give a semiparametric example.

\begin{example}\label{exmp:3}
Consider   model (f)  and    assume that $f(z,u,w\mid x)$  is a  normal density for all $x$;
then $f(w\mid z,x)$  is a normal density function.
If one has available $\pr(y\mid z,x)$ and $f(w\mid z,x)=1/\sigma(x)\phi\{(w-\beta_0(x)-\beta_1(x)z)/\sigma(x)\}$  from observed variables, with $\phi$  the standard normal density function and $\beta_1(x)\neq 0$ for all $x$, then \eqref{eq:intg} has a unique solution  
\[h(w,x,y)=\frac{1}{2\pi}\int_{-\infty}^{+\infty}\exp\left\{\frac{{\rm i}vw}{\sigma(x)}\right\}\frac{h_2(v,x,y)}{h_1(v)} dv,\]
where ${\rm i}=(-1)^{1/2}$, $h_1(v)$ and $h_2(v,x,y)$ are Fourier  transforms of  $\phi$  and $\pr(y\mid z,x)$ respectively:  $h_1(v)=\int_{-\infty}^{+\infty}\exp(-{\rm i}vz)\phi(z)dz$, and 
\[h_2(v,x,y)=\frac{\beta_1(x)}{\sigma(x)}\int_{-\infty}^{+\infty} \exp\left\{-{\rm i}v\frac{\beta_0(x)+\beta_1(x)z}{\sigma(x)}\right\} \pr(y\mid z,x)dz.\]
After obtaining $h(w,x,y)$, we can further identify $\pr\{y\mid {\rm do}(x)\}$ from \eqref{eq:ce2}.
\end{example}
Example  \ref{exmp:3} allows $X$ and $Y$ to follow an arbitrary distribution, not necessarily  normal, so the model is semiparametric and thus  is a generalization of the joint normal model for $(X,Y,U,W,Z)$ considered  by \cite{kuroki2014measurement}.
Under  the model   of \cite{kuroki2014measurement},  Example \ref{exmp:3} can be  strengthened: one only needs to  implement  linear regression to obtain $f(y\mid z,x)\thicksim N(\alpha_0+\alpha_1z+\alpha_2x,\sigma_1^2)$ and  $f(w\mid z,x)\thicksim N(\beta_0+\beta_1z+\beta_2x,\sigma_2^2)$, then $\pr\{y\mid \text{do}(x)\} \thicksim N(\gamma_0+\gamma_1x,\sigma^2)$ with $\gamma_1=\alpha_2 - \alpha_1\beta_2/\beta_1$ and  $(\gamma_0,\sigma^2)$ presented in the Supplementary Material. One can  verify that $\gamma_1=\partial E(y\mid u,x)/\partial x$, which is     consistent with the result of  \cite{kuroki2014measurement} obtained by an analysis of  variance approach that entails   normality of $(X,Y)$. 

 

\section{Hypothesis testing with proxy variables}
Considering  a categorical confounder $U$ with $k$ levels,  condition (i) implicitly requires that both  $Z$ and $W$ have at least as many categories as $U$, otherwise,   the causal effect is  in general not identifiable. Nevertheless, as we elaborate below, when one of the proxies has fewer categories than $U$, we  use such proxy variables to test the  causal null hypothesis
$\mathbb H_{0}:\ X\ind Y\mid U$  in model (f),   then we generalize the results to  other diagrams. 
The null hypothesis  means that $X$ has no causal effect on $Y$ at any level  of $U$, which is equivalent to  $\pr(y\mid u)=\pr(y\mid u,x)$ for all $x,y$ and $u$.  Thus, rejection of the null hypothesis is  evidence in favor of causation between $X$ and $Y$.  Because $U$ is not observed, we cannot directly check the divergence  between $\pr(y\mid u,x)$ and $\pr(y\mid u)$. 
Nevertheless, we propose a  measure for  $\pr(y\mid u,x)-\pr(y\mid u)$ based on proxy variables of $U$.    We assume
\begin{enumerate}
\item[(iv)] $X,Z$  and $W$ have $i,j$ and $k$ categories respectively, with  $i j\geq k+1$;   and  the matrix $Q=\{P(W \mid Z, x_1),\ldots,P(W \mid Z, x_i)\}$ has full row rank.
\end{enumerate}
Under (iv),  $P(W\mid U)$ is invertible \citep[Corollary 5.4]{banerjee2014}, and \eqref{eq:yzt2} still holds: $P(y\mid Z, x)   =    P(y\mid U, x)  P(W\mid U) ^{-1} P(W\mid Z, x)$ for all $x$.  
Denoting   $q_y=\{P(y\mid Z, x_1),\ldots,P(y\mid Z,x_i)\}^{\rm T}$, we have the decomposition
\begin{eqnarray}\label{eq:vari}
q_y^{\rm T}\ & = &  P(y\mid U)  P(W\mid U) ^{-1}Q + \{P(y\mid U,x)-P(y\mid U)\}P(W\mid U) ^{-1}Q.
\end{eqnarray}
For fixed $y$, as $x$ varies, \eqref{eq:vari}  reveals two separate sources of variability of  $P(y\mid Z,x)$: $P(W\mid Z,x)$ and $P(y\mid U,x)$. If   $\mathbb H_0$ holds, $P(W\mid Z,x)$ is  the only source  of variability because $P(y\mid U,x)=P(y\mid U)$.
Thus, we can test $\mathbb H_0$ by checking whether $P(W\mid Z,x)$ explains away the  variability of $P(y\mid Z,x)$. 

Suppose we  have available estimators $(\widehat q_y,\widehat Q)$ that satisfy
\begin{eqnarray}
&n^{1/2}(\widehat q_y  - q_y )\rightarrow  N(0,\Sigma_y) \text{ in distribution,}  \label{eq:est1}\\
&\widehat Q  \rightarrow Q \text{ and } \widehat\Sigma_y \rightarrow \Sigma_y \text{ in probability, with  $\widehat \Sigma_y, \Sigma_y$  positive-definite.} \label{eq:est2}
\end{eqnarray}
Letting $I$ denote the identity matrix and 
\begin{eqnarray*}
\xi_y=\{I - \widehat \Sigma_y^{-1/2}\widehat Q^{\rm T}(\widehat Q\widehat \Sigma_y^{-1} \widehat Q^{\rm T})^{-1}\widehat Q\widehat \Sigma_y^{-1/2}\} \widehat \Sigma_y^{-1/2} \widehat q_y,
\end{eqnarray*}
then $\xi_y$ is  the least-square residual of regressing $\widehat \Sigma_y^{-1/2} \widehat q_y$ on $\widehat \Sigma_y^{-1/2} \widehat Q$, and thus measures the residual variability of $P(y\mid Z,x)$ not explained by $P(W\mid Z,x)$.
Therefore, we can check how far away $\xi$ is from zero to assess whether $\mathbb H_0$ is correct by using the  test statistic 
$T_y=n\xi_y^{\rm T}\xi_y$.

\begin{theorem}\label{thm:tst}
Assuming model (f), conditions (iv) and \eqref{eq:est1}--\eqref{eq:est2}, if $\mathbb H_0$ is correct,  then 
$n^{1/2} \xi_y  \rightarrow  N(0,\Omega_y)$ in distribution, with $\Omega_y=I-\Sigma_y^{-1/2}Q^{\rm T}(Q\Sigma_y^{-1}Q^{\rm T})^{-1}Q\Sigma_y^{-1/2} $ of rank $r=ij-k$, 
 and $T_y \rightarrow \chi_{r} ^2$ in distribution.
\end{theorem}
From Theorem \ref{thm:tst},  $ \xi_y$ asymptotically follows a  degenerate multivariate    normal distribution, further explaining why  $ij \geq k+1$ is required.
Given a significance level $\alpha$, one  can reject $\mathbb H_0$ as long as  $T_y$ exceeds  the $(1-\alpha)$th quantile of $\chi_{r}^2$, which guarantees  a type I error 
no larger than $\alpha$ asymptotically.  
Theorem \ref{thm:tst} concerns only one level of the outcome, but in the Supplementary Material we  extend  to  aggregating   all levels of a categorical $Y$.
For a categorical outcome, \eqref{eq:est1}--\eqref{eq:est2}  can be achieved for instance with  empirical probability mass functions  $\widehat \pr(w\mid z,x)$ and $\widehat \pr(y\mid z,x)$.
For a continuous one,   \eqref{eq:est1}--\eqref{eq:est2} are   generally not feasible  and discretization  is required; however,  in many situations where the average causal effect is of interest,  
one can use   $q=\{E(Y\mid Z, x_1),\ldots,E(Y\mid Z,x_i)\}^{\rm T}$ in construction of the test statistic and  perform the test on the mean scale.

The proposed testing strategy  is readily generalized to accommodate polytomous $W$ with more than $k$ levels by using an appropriate coarsening of $W$ to construct the test statistic.
The proposed test for  model (f)   applies to (d)--(e).
For models (b)--(c), we can test $\mathbb H_0$  by treating 
one of the proxies as a constant, and to guarantee $ij\geq k+1$, we require $Y$ and $X$ to  have more categories than $k$, respectively. 
The result equally applies to model (a), which is  a special case of (b) and (c). 
Simulations  confirm that  our  testing strategy performs reasonably well for 
a moderate sample size, with  type I error approximating  the nominal level  when $\mathbb H_0$ holds and statistical power increasing toward unity when $\mathbb H_0$ does not hold.


\section{Discussion}
Identifiability  of $\pr(w\mid u)$ in models  (d)--(e) reflects the well-known fact in latent class analysis  that  the  joint model is identified with at least  three independent  proxies for the latent factor  \citep{kruskal1976,goodman1974,allman2009}. However,  our analysis  for  model (f)  highlights that  certain parameters of interest such as the causal effect,   is still identifiable  with only two independent proxies, even though the latent class model is not completely identified. 
This work also has promising application in  observational studies when negative controls are available \citep{lipsitch2010negative,gagnon2012using,sofer2016,miao2017invited},
in which case, a negative control outcome that is not causally affected by the treatment, and a negative control exposure that does not causally affect the primary outcome,  may serve as  proxies for the confounder. 
The proposed methods can accommodate observed covariates by  incorporating them into all conditional densities and marginalizing over them to obtain the causal effect.
The identification results can be extended by considering multiple   confounders $U=(U_1,\ldots,U_l)$ with multiple proxies $Z=(Z_1,\ldots, Z_l)$ and $W=(W_1,\ldots, W_l)$ such that  model (f)  holds,  i.e.,  $W\ind (Z,X)\mid U$, $Z\ind Y\mid (U,X)$,  in which case,  Theorem \ref{thm:idn}  still applies. 
For the continuous confounder case,  estimation in parametric models, for instance in  normal models, is straightforward by linear regression. However, estimation  is very challenging in nonparametric models, as it  requires solving an integral equation, which we will study elsewhere.

\section*{Acknowledgement}
The work is partially supported by the China Scholarship Council and the National Institute of Health. The authors are grateful to the editor and  three referees   for their helpful comments.

\section*{Supplementary material}
Supplementary material available at \Bka\  online includes   examples,  proofs, discussion on the solution to \eqref{eq:intg},   details for Example \ref{exmp:3}, and  simulations for the testing strategy.

\section*{Appendix}
We use the singular value decomposition \citep[Theorem 2.41]{carrasco2007}  of compact operators to characterize conditions for existence of a solution to \eqref{eq:intg}.
Letting $L^2\{F(t)\}$  denote the space of all square integrable functions of  $t$ with respect to a cumulative distribution function $F(t)$, which is  a Hilbert space with  inner product 
$\langle g, h\rangle= \int_{-\infty}^{+\infty} g(t) h(t) dF(t) $,
letting $K_x$ denote the conditional expectation operator: $L^2\{F(w\mid x)\}\longmapsto L^2\{F(z\mid x)\}$, $K_xh=E\{h(w)\mid z,x\}$ for  $h \in L^2\{F(w\mid x)\}$,
and letting  $(\lambda_{x,n},\varphi_{x,n},\psi_{x,n})_{n=1}^{+\infty}$  denote a singular value decomposition  of  $K_x$, 
we assume the following regularity conditions:
\begin{enumerate}
\item[(v)] $\int_{-\infty}^{+\infty}\int_{-\infty}^{+\infty} f(w\mid z,x)f(z\mid w,x)dwdz < +\infty$,
\item[(vi)]  $\int_{-\infty}^{+\infty} \pr^2(y\mid z,x)f(z\mid x)dz < +\infty$,
\item[(vii)]  $\sum_{n=1}^{+\infty} \lambda_{x,n}^{-2} |\langle \pr(y\mid z,x), \psi_{x,n} \rangle|^2 <+\infty$,
\end{enumerate}
then we have the following proposition.
\begin{proposition}\label{prop:1}
Given $f(w\mid z,x)$ and $\pr(y\mid z,x)$, the solution to \eqref{eq:intg} must exist if conditions (iii) and (v)--(vii) hold together.
\end{proposition}

\bibliographystyle{biometrika}
\bibliography{CausalMissing}
\label{lastpage}

\end{document}